\documentclass[]{spie}

\usepackage{graphicx}

\title{Deflection of Light and Shapiro Delay: \\An Equivalent Medium Theory Approach}

\author{Sina Ataollah Khorasani
\skiplinehalf
School of Electrical Engineering, Sharif University of Technology, Tehran, Iran
}

\authorinfo{\noindent Further author information: (Send correspondence to S.K.)\\S.K.: E-mail: khorasani@sina.sharif.edu}

\begin{document}
\maketitle

\begin{abstract}
We discuss the deflection of light and Shapiro delay under the influence of gravity as described by Schwarzschild metric. We obtain an exact expression based on the coordinate velocity, as first set forth by Einstein, and present a discussion on the effect of velocity anisotropy. We conclude that the anisotropy in the coordinate velocity, as the velocity apparent to a distant observer, gives rise to a third order error in the deflection angle, so that the practical astronomical observations from gravitational lensing data remain inconclusive on the anisotropy. However, measurement of Shapiro delay provides a fairly convenient way to determine whether the spacetime is optically anisotropic for a distant observer or not. We calculate the Shapiro delay for a round trip path between Earth and Venus and observe excellent agreement to two experimentally reported values measured during a time span of six months in 1967, without any need to extra fitting parameters. This is while the expected delay obtained from an isotropic light velocity as described by Einstein's model suffers from much larger errors under similar conditions. This article illustrates the usefulness of the equivalent medium theory in understanding of general theory of relativity.
\end{abstract}


\section{INTRODUCTION}{\label{Sec1}}

It is widely known that Schwarzschild metric is the only exact solution of the Einstein field equations under spherical symmetry, which is given by the metric \cite{1,2}

\begin{equation}{\label{eq1}}
c_0^2d\tau^2=c_0^2(1-\frac{r_s}{r})dt^2-\frac{dr^2}{1-\frac{r_s}{r}}-r^2(d\phi^2+\sin^2\phi d\theta^2),
\end{equation}

\noindent
where $\tau$ is the proper time, $c_0$ is the speed of light in flat space, $(r,\theta,\phi)$ constitute the spherical polar coordinates, and $r_s$ is the Schwarzschild radius of the star. Initially in 1911, Einstein mentioned that the light velocity arising from such a spherically symmetric gravity should be modified as \cite{1}

\begin{equation}{\label{eq2}}
c(\textbf{r})=c_0\left[1+\Phi(r)\right],
\end{equation}

\noindent
in which $\Phi(r)=-\frac{r_s}{2r}$ is the gravitational potential. He furthermore took notice of this radial dependence of light velocity to propose the possibility of deflection of light while passing across a massive object according to the angle

\begin{equation}{\label{eq3}}
\alpha=\frac{r_s}{R},
\end{equation}

\noindent
with $R$ being the closest distance of approach of the light ray with respect to the center of the massive object. But later in 1916, Einstein noticed a missing correction factor of $2$ in (\ref{eq2}) as

\begin{equation}{\label{eq2c}}
c(\textbf{r})=c_0\left[1+2\Phi(r)\right],
\end{equation}

\noindent
resulting in

\begin{equation}{\label{eq4}}
\alpha=2\frac{r_s}{R}.
\end{equation}

\noindent
Famously, this result was confirmed in the 1919 experiment of Sir Arthur S. Eddington \cite{4}.

Perhaps, Eddington was the first to point out explicitly that ``For the original coordinates of (\ref{eq1}) the velocity of light is not the same for the radial and transverse directions'' \cite{5}. To resolve this difficulty, he proposed the change of coordinates using the transformation

\begin{equation}{\label{eq5}}
r=(1+\frac{r_s}{4\rho})^2\rho,
\end{equation}

\noindent
which takes advantage of the non-physical radial coordinate $\rho$, and subsequently results in the isotropic metric

\begin{equation}{\label{eq6}}
c_0^2d\tau^2=c_0^2\frac{(1-\frac{r_s}{4\rho})^2}{(1+\frac{r_s}{4\rho})^2}dt^2-(1+\frac{r_s}{4\rho})^4(d\rho^2+\rho^2d\phi^2+\rho^2\sin^2\phi d\theta^2).
\end{equation}

\noindent
It is readily possible to calculate the resulting \textit{coordinate velocity} of light in this frame to be

\begin{equation}{\label{eq7}}
c(\textbf{r})=c_0\frac{1-\frac{r_s}{4\rho}}{(1+\frac{r_s}{4\rho})^3}\approx c_0\left[1+2\Phi(r)\right],
\end{equation}

\noindent
where the approximation $r\approx\rho$ holds in the limit of $\rho>>r_s$, according to (\ref{eq5}). The light velocity (\ref{eq7}) as measured from a distant observer is apparently isotropic, while a local observer regardless of the choice of coordinates would always see a constant and direction-independent speed of $c_0$, which is of course also independent of the choice of coordinates. This also follows the principle of equivalency, which states that any accelerated observer would always see a flat space to a first-order of accuracy.

As a matter of fact, it is easy to observe that there is a maximum anisotropy in the coordinate speed of light as suggested by (\ref{eq1}) equal to $|\Phi(R)|c_0$, corresponding to the difference between radial and transverse speeds, the exact relation of which is found through the equivalent medium theory to be \cite{6,7}

\begin{equation}{\label{eq8}}
c(\textbf{r,k})=c_0\sqrt{1+2\Phi(r)}\sqrt{1+2\Phi(r)\cos^2\Psi(\textbf{r,k})}.
\end{equation}

\noindent
Here, $\Psi(\textbf{r,k})=\textrm{arccos}(\hat{r}\cdot\hat{k})$ is the angle made by the unit radial $\hat{r}=\frac{\textbf{r}}{r}$ and unit propagation vectors $\hat{k}=\frac{\textbf{k}}{k}$, and hence is anisotropic as seen by a distant observer. Interestingly, in the limit of large radius $r$, (\ref{eq8}) simplifies to

\begin{equation}{\label{eq9}}
c(\textbf{r,k})\approx c_0\{1+\left[1+\cos^2\Psi(\textbf{r,k})\right]\Phi(r)\},
\end{equation}

\noindent
which establishes the fact that Einstein's correction coefficient of $2$ indeed varies between $1$ and $2$ depending on the angle of propagation. Evidently, this value remains close to 2 as long as the light ray is at a position $r$ away from its closest distance of approach $R$; it only for the region sufficiently close to the closest distance of approach, appreciably decreases to achieve a minimum of $1$, exactly at $r=R$. The above result (\ref{eq9}), which is obtained directly through the equivalent medium theory \cite{6,7}, has been also deduced by Eckstein \cite{7a} through geometric considerations. It is furthermore noticed that the corresponding anisotropy in the light velocity due to the gravitational field of sun and the black hole at the center of milky way galaxy could respectively reach $9\times10^{-9}$ \cite{6,7}, and $5\times10^{-7}$ according to Mizushima \cite{8}.

In 1959, Yilmaz \cite{9} proposed an experiment which put the equivalence principle of the general relativity to the test. He suggested to set up an optical interferometer, looking for any such possible local anisotropy arising from the gravity of Sun. Ten years later, Shamir and Fox published a paper \cite{10} reporting null results within an accuracy of $3\times10^{-11}$.

The purpose of this manuscript is two fold. We examine the deflection of light as well as Shapiro delay as plausible methods to detect the possible anisotropy in the light speed:

\begin{itemize}
\item
First, we show that the exact deflection angle arising from an anisotropic light velocity (\ref{eq8}) could meaningfully differ from the value as predicted from the isotropic velocity (\ref{eq2c}). For this purpose, we first derive an exact expression for the deflection angle as suggested from (\ref{eq2c}). This expression agrees with the more accurate expression reported recently by Virbhadra \cite{11,12} to the second-order. Then we numerically evaluate the exact angle of deflection based on (\ref{eq8}) to notice an appreciable difference when $R\leq 5r_s$. We discuss that for all practical reasons the deflection of light is not a good means to decide on the anisotropy of equivalent medium space, since the expected difference is vanishingly small when $R >> r_s$.

\item
Secondly, we consider the Shapiro delay. We calculate this delay based on anisotropic description of space and observe an excellent agreement to the experimental data, without any need to an external parameter. This is while either of the 1911 or 1916 Einstein's isotropic expressions suffer from much larger errors.
\end{itemize}

\section{DEFLECTION OF LIGHT}
\subsection{Isotropic Coordinate Velocity}

The significant difference between the concepts of \textit{coordinate velocity} and \textit{local velocity} of light has been a matter of debate in numerous works, which is summarized adequately by Longhi \cite{13}. More recently, it has been stated by Jensen \cite{14} that the Schwarzschild metric could violate the weak principle of equivalence, which is here understood to be a misinterpretation of the concept of coordinate velocity. Similarly, it has been argued by the author \cite{7,15} that whether local interferometry could ultimately detect gravitational waves or not; this is here left as the subject of a separate study.

\begin{figure}
\begin{center}
\includegraphics[width=12cm]{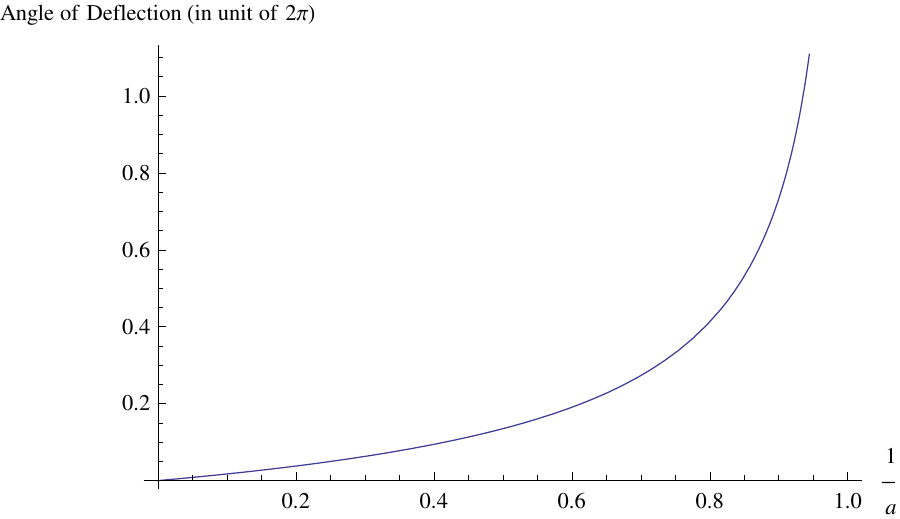}%
\caption{Exact angle of deflection versus inverse normalized parameter $\frac{1}{a}$.\label{Fig1}}
\end{center}
\end{figure}

In order to calculate the angle of deflection, we follow the integral method originally used by Einstein \cite{3}. For a light ray moving on the plane $(x_1,x_3)$ subject to the gravitational field of a massive object located at the origin, the deflection angle $\alpha$ can be found as \cite{3}

\begin{equation}{\label{eq10}}
\alpha=\int_{-\infty}^{+\infty}\frac{1}{c(\textbf{r})}\frac{\partial c(\textbf{r})}{\partial x_1} dx_3,
\end{equation}

\noindent
with $\textbf{r}=(x_1,x_2,x_3)$ and $x_2\equiv0$. Upon plugging (\ref{eq4}) in (\ref{eq10}) we get after significant algebra

\begin{equation}{\label{eq11}}
\alpha(a)=a\frac{\pi-\textrm{Arctan}\sqrt{a^2-1}}{\sqrt{a^2-1}}-\frac{\pi}{2},
\end{equation}

\noindent
in which the definition $a=\frac{R}{2r_s}$ is adopted. In the limit of $R\rightarrow 2r_s$ the angle $\alpha$ approaches infinity, which can be interpreted as infinite rotation at $R=2r_s$. Since the photon sphere is known to be located at $R=\frac{3}{2}r_s$, the expression (\ref{eq11}) might be regarded as invalid for $a$ close to unity. This behavior is illustrated in Fig. \ref{Fig1}.

When $a>1$, this expression may be expanded in the powers of $\frac{1}{a}$ to obtain

\begin{equation}{\label{eq12}}
\alpha(a)\approx\frac{1}{a} + \frac{\pi}{4}\frac{1}{a^2}+\frac{2}{3}\frac{1}{a^3}+\frac{3\pi}{16} \frac{1}{a^4}+\frac{8}{15}\frac{1}{a^5}.
\end{equation}

\noindent
This agrees to the Einstein's result (\ref{eq4}) to the first order. In contrast, Virbhadra's expression reads \cite{11,12}

\begin{equation}{\label{eq13}}
\alpha(a)\approx\frac{1}{a} + \frac{1}{4}(\frac{15\pi}{16}-1)\frac{1}{a^2},
\end{equation}

\noindent
which has been obtained from direct integration of an extended expression by Weinberg \cite{16} to the case of the most general static spherically symmetric asymptotically flat universe.

\subsection{Anisotropic Coordinate Velocity}

Now, we proceed to calculate the deflection angle as caused by using the exact anisotropic coordinate velocity (\ref{eq8}). We may use

\begin{equation}{\label{eq14}}
\alpha\approx\int_{-\infty}^{+\infty}\frac{1}{c(\textbf{r,k})}\frac{\partial c(\textbf{r,k})}{\partial x_1} dx_3.
\end{equation}

\noindent
The difficulty in evaluating the integral is that the propagation vector $\textbf{k}$ changes its direction along the light trajectory. Hence, we first rewrite (\ref{eq14}) as

\begin{equation}{\label{eq15}}
\alpha(\textbf{r})\approx2\int_{0}^{x_3}\frac{1}{c(\textbf{r}^\prime,\textbf{k}^\prime)}\frac{\partial c(\textbf{r}^\prime,\textbf{k}^\prime)}{\partial x^\prime_1} dx^\prime_3.
\end{equation}

\noindent
Upon differentiating with respect to $x_3$ from both sides we arrive at the first order nonlinear differential equation

\begin{equation}{\label{eq16}}
\frac{d}{d x_3}\alpha(\textbf{r})\approx2\frac{1}{c(\textbf{r,k})}\frac{\partial c(\textbf{r,k})}{\partial x_1},
\end{equation}

\noindent
in which we have

\begin{equation}{\label{eq17}}
\hat{k}=\cos\alpha(\textbf{r})\hat{x}_1+\sin\alpha(\textbf{r})\hat{x}_3.
\end{equation}

\noindent
Now, using the initial condition $\alpha(R,0,0)=0$ we can obtain the deflection angle as $\alpha=\lim_{z\rightarrow\infty}\alpha(R,0,z)$, which can be easily done by means of accurate numerical integration methods.

\subsection{Comparison}

\begin{figure}
\begin{center}
\includegraphics[width=12cm]{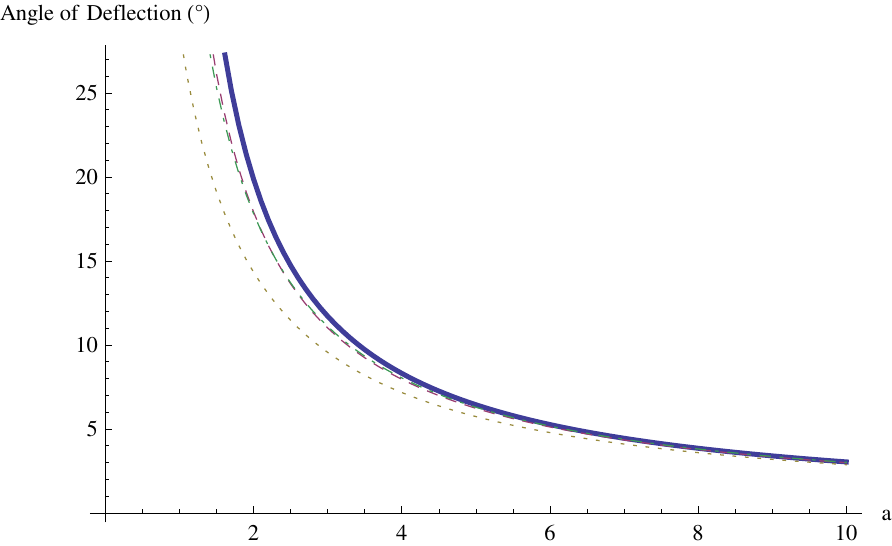}%
\caption{Angle of deflection from various expressions versus normalized parameter $a$. \label{Fig2}}
\end{center}
\end{figure}

The variation of deflection angle $\alpha$ versus the closest distance of approach $R$ is here plotted as shown in Fig. \ref{Fig2}, based on the four expressions. Dotted, dashed, and dot-dashed curves respectively represent the Einstein (\ref{eq4}), Virbhadra (\ref{eq13}), and our expression (\ref{eq11}). In contrast, the solid heavy curve represents the angle of deflection (\ref{eq14}) based on anisotropic coordinate velocity.

In general, the difference between the deflection angles from anisotropic and isotropic coordinate velocities is less than 1\% when $a>5$. So, observation of angles of deflection may not be an efficient way to determine whether the space behaves optically anisotropic or not.

\section{SHAPIRO DELAY}

Calculation of Shapiro delay \cite{17,18,19} is a surprisingly convenient method, as we here show, to determine whether the space looks optically isotropic for a distant observer or not. In 1964, Irwin I. Shapiro proposed \cite{17} an interesting measurement to validate the general theory of relativity, through observation of round trip delay of a radio signal traveling between Earth and Venus, while the radio beam grazes the solar limb. Since all the existing relations for light velocity are independent of frequency, the same equations can be applied to radio waves. Under the influence of Sun's gravity, the speed of the electromagnetic wave would decrease. As a result of this slowness, a time-difference can be detected by subtracting the round trip delay of a radio beam propagating at the constant speed of $c_0$, from the actual measured value, which can be attributed to the influence of the gravitational field. Based on an isotropic light velocity as suggested by Einstein in 1916 (\ref{eq2c}), Shapiro's predicted value turns out to be about $200\mu\textrm{sec}$ \cite{17}, while using the Einstein's 1911 result (\ref{eq2}) one may expect half as much, or roughly $100\mu\textrm{sec}$. Soon after, Ross and Schiff \cite{20} noticed the dependence of calculations on the particular choice of coordinates of either (\ref{eq1}) or (\ref{eq8}), and suggested the correct transformation to eliminate the apparent coordinate dependence. More recently, Shapiro delay has been used to accurately measure the mass of a Neutron star \cite{21}.

\begin{figure}
\begin{center}
\includegraphics[width=12cm]{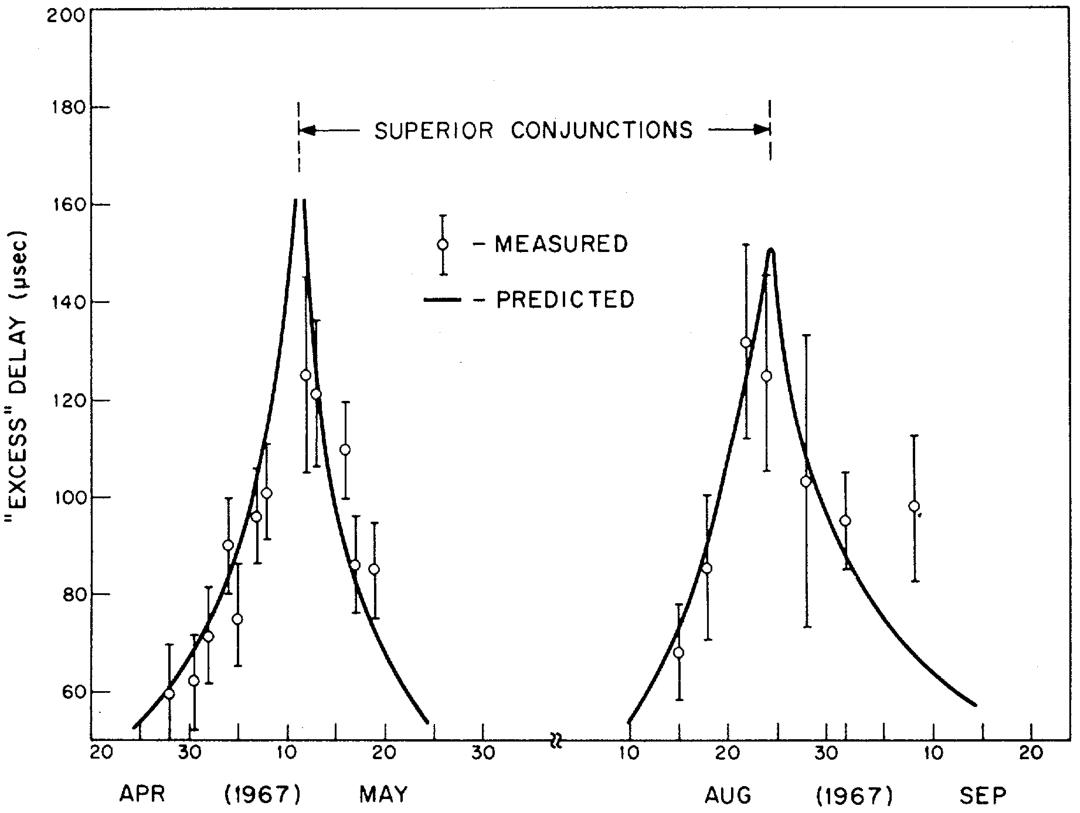}
\caption{Shapiro delay \cite{18}. \label{Fig3}}
\end{center}
\end{figure}

Two measurements actually took place in May and September 1967 \cite{18}, respectively giving rise to the values of $160\mu\textrm{sec}$ and $150\mu\textrm{sec}$, as illustrated in Fig. \ref{Fig3}. Shapiro's observations \cite{18} matched the theory within the relatively large error of $\pm20$\%, while using a generalized metric with $\frac{1}{2}(1+\gamma)=0.9\pm0.2$, instead of the Schwarzschild's with $\frac{1}{2}(1+\gamma)=1$. It has been more recently shown through later astronomical observations of gravitational lensing \cite{19} by Cassini spacecraft, that this expression $\frac{1}{2}(1+\gamma)$ should be very close to unity, indeed within an error margin of only $\pm10^{-5}$.

It is fairly easy to make an estimate of the Shapiro delay based on the anisotropic coordinate velocity (\ref{eq6}). We may here write down

\begin{equation}{\label{eq18}}
\Delta t\approx2\int_{-a_T}^{+a_R}\left[\frac{1}{c_0}-\frac{1}{c(\textbf{r,k})}\right] dx_3,
\end{equation}

\noindent
in which $a_T$ and $a_R$ are respectively the orbiting radii of Earth and Venus around the Sun, being equal to $1.496\times10^{11}\textrm{m}$ and $1.0829\times10^{11}\textrm{m}$. The closest distance of approach $R$ is taken to be the solar radius $6.96\times10^8{m}$ having a Schwarzschild radius of approximately $3\times10^3\textrm{m}$. Numerical evaluation of the integral (\ref{eq18}) gives rise to the value of $148.342\mu\textrm{sec}$, in surprisingly good agreement with the observed values of $150-160\mu\textrm{sec}$. This is while using the 1911 (\ref{eq2}) and 1916 (\ref{eq2c}) expressions give respectively the values of $118.113\mu\textrm{sec}$ and $236.226\mu\textrm{sec}$. It should be added here that in this calculation, we have assumed circular orbits for Venus and Earth, while their actual orbits have eccentricities of respectively 0.68\% and 1.7\%. This agreement also narrows the experimentally determined value of $\gamma=1$ through radio Shapiro delay measurement to less than 1\%, thus reaching to the best of similar observations so far \cite{19}. It is noticeable that Eckstein \cite{7a} has also used (\ref{eq9}) to obtain a confirming expression to that of the generally accepted formula in \cite{19}.

This may provide a very good basis for the understanding of how coordinate and local velocities of light differ in practice. Actually, the coordinate velocity becomes meaningful in the framework of equivalent medium theory, and we have shown the accuracy of this approach as an alternative calculation method. The calculations take place on a standard Minkowskian or Cartesian frame of coordinates, instead of the curved space. Moreover, this reasonable agreement between our calculation and observed data reveals the good accuracy of the general theory of relativity, without any need for any externally inserted fitting parameters. However, this observation relies on the fact that the velocity of light as appears to a distant observer is actually anisotropic under the influence of gravitational field. In short, we may deduce that the coordinate velocity, or the apparent velocity to a distant observer who is outside a gravitational field, actually corresponds to the same quantity obtained from the standard equivalent medium theory.

\section{CONCLUSIONS}

In this paper, we discussed the possible influence of gravity-induced anisotropy on the propagation of optical waves within the framework of equivalent medium theory. We discussed two plausible approaches to examine the physical influence of this anisotropy through measurement of deflection of light and also the so-called Shapiro delay. We concluded that for practical reasons, the deflection of light is not sufficiently accurate to distinguish between isotropic and anisotropic coordinate velocities of light in the equivalent medium framework. On the other hand, measurement of the Shapiro delay turns out to be an ideal method to test the anisotropy of coordinate light velocity, as it appears to a distant observer. We found a fairly good agreement between the calculated value based on an anisotropic light velocity and the experimentally observed value, being within 1\% and 7\% of the observed values, belonging to two separate measurements.

\end{document}